\documentclass[journal]{IEEEtran}
\usepackage{graphicx,epsfig,color,graphics}
\newtheorem{theorem}{Theorem}
\newtheorem{lemma}{Lemma}

\newtheorem{problem}{Problem}
\newtheorem{cor}{Corollary}

\newtheorem{observation}{Observation}

\begin{document}


\title{On the Complexity of finding Stopping Distance in Tanner Graphs 
\thanks{A preliminary version of this paper was presented at CISS 2006,
March 2006}}


\author{
  \authorblockN{K. Murali Krishnan , Priti Shankar \\
      \authorblockA{Department of Computer Science and Automation \\
                    Indian Institute of Science, Bangalore - 560012, India.\\
                    Email: \{kmurali,priti\}@csa.iisc.ernet.in}}}

\maketitle

\begin{abstract}
Two decision problems related to the computation of stopping sets in 
Tanner graphs are shown to be NP-complete. NP-hardness of the  
problem of computing the stopping distance of a Tanner graph follows
as a consequence.
\end{abstract}

\IEEEpeerreviewmaketitle

\section{Introduction}
\label{intro}

Stopping sets were introduced in \cite{Di} for the analysis of erasure
decoding of LDPC codes. It was shown that
the iterative decoder fails to decode to a codeword if and only
if the set of erasure positions is a superset of some stopping set in the
Tanner graph \cite{Tan} used in decoding.
Considerable analysis has been carried out on the 
distribution of stopping set sizes in LDPC code ensembles, giving valuable
insight into the asymptotic performance 
of message-passing decoding on LDPC ensembles --- see for example  
\cite{Di2,Or}.
  Since small stopping sets are directly responsible for 
poor performance of iterative decoding algorithms, it is of interest to 
determine the size of the smallest stopping set in a Tanner graph,
called the {\em stopping distance} of the graph.  Construction of 
codes for which there are Tanner graphs that do not contain small
stopping sets has been studied --- see for example \cite{Tian,Adi}.  
The stopping distance of the
graph, is of interest as it gives the minimum number of erasures
that can cause iterative decoding to fail.  

The relationship between stopping distance
and other graph parameters like girth has been explored in \cite{Or2}
where it is shown that large girth implies high stopping distance.
Pishro-Nik and Fekri \cite{Nik} showed that by adding a suitable number 
of parity checks the stopping distance of a Tanner graph for a code
can be increased to the maximum possible, viz., the minimum distance of 
the code.  Schwartz and Vardy \cite{Var} defines the {\em stopping
redundancy} of a code as the minimum number of rows in a parity check
matrix for the code such that the stopping distance of the corresponding
Tanner graph is equal to the minimum distance of the code 
and proves some bounds on the stopping redundancy for
various classes of codes.  Further investigations on stopping redundancy 
may be found in  \cite{Sgl}.

In this correspondence, we show that the computational problems of 
determining whether a given Tanner graph has a stopping set of a given size 
or at most a given size are NP-complete.  
These are shown by reductions from the well known NP-complete problems of 
determining whether a given graph contains a vertex cover of a given size 
(respectively at most a given size) to the above problems.   NP-hardness
of the problem of finding the stopping distance of a Tanner graph 
follows as a consequence of the latter result.

\section{Background}
\label{background}

Given a parity check matrix $H = [h_{ij}] \in GF(2)^{(n-k)\times n}$,
$1\leq k\leq n$ for an $(n,k)$  binary linear code, the 
Tanner graph is  the undirected bipartite graph $G=(L,R,E)$ 
where $L=\{ x_{i}, 1\leq i\leq n \}$, $R=\{ c_{j}, 1\leq j\leq n-k\}$
and $E=\{ (x_{i},c_{j}): h_{ji}=1, 1\leq i\leq n, 1\leq j\leq n-k\}$.
The set $L$ corresponds to the set of codeword elements 
and $R$ corresponds to the set of parity checks.
We refer to the set $L$ and $R$ as the set of left and right vertices 
respectively.  For $S\subseteq L\cup R$, we define 
$N(S)=\{ y : (x,y)\in E, x\in S\} $. $S\subseteq L$ is a 
{\em stopping set} if for all $c_{j}\in N(S)$,
$|N(\{ c_{j}\} )\cap S|\geq 2$ ie., every vertex connected to some vertex
in a stopping set must have at least two neighbours in the the stopping set.
The {\em stopping distance} of a Tanner graph is the size of
the smallest stopping set in the graph.  We define two
decision problems concerning stopping sets:

\begin{problem}
STOPPING SET:  Given a Tanner graph $G$ and positive integer $t$,
does $G$ have a stopping set of size $t$.
\end{problem}

\begin{problem}
STOPPING DISTANCE:  Given a Tanner graph $G$ and positive integer $t$,
does $G$ have a stopping set of size at most $t$.
\end{problem}

Note that the corresponding decision problems arising out of the
problem of finding the minimum distance of a code
were shown to be NP-complete in \cite{Mac} and \cite{Var2}.

It is clear that if either STOPPING SET or STOPPING DISTANCE can 
be solved in polynomial
time, then evoking the algorithm at most $|L|$ times, the problem of
actually finding the stopping distance of a Tanner graph can be
solved.   Conversely, if there is a polynomial time algorithm for
finding the stopping distance of a given Tanner graph $G$, then
we can use the algorithm to solve STOPPING DISTANCE since $G$ has
stopping distance less than or equal to $t$ if and only if $G$ contains
is a stopping set of size less than or equal to $t$.  Note
that it is not immediately clear how to solve STOPPING SET in polynomial
time even if a polynomial time algorithm for computing the stopping 
distance of a Tanner graph is known.

The notion of NP-completeness was introduced in \cite{Cook}, and
is well established in the computer science literature for the analysis of
the computational complexity of problems (see \cite{Cor, Gar} for
a detailed account).  Typically, a problem is posed
as a decision problem, i.e., one where the solution consists
of answering it with a {\em yes} or a {\em no}.  
All inputs for which the answer
is a {\em yes} from a set.  We identify this set with the problem.
A decision problem $A$ belongs
to the class NP if there exists a polynomial time algorithm $\Pi $
such that, for all $x\in A$, there exists a string $y$ (called a
{\em certificate} for membership of $x$ in $A$), with $|y|$ polynomially
bounded in $|x|$, such that $\Pi$ accepts $(x,y)$, whereas, for all
$x\notin A$, $\Pi$ rejects $(x,y)$ for any string $y$ presented to the
algorithm.  In other words, problems in NP are precisely those for which 
membership verification is polynomially solvable.  
We say a decision problem $A$ is {\em polynomial time 
many-one reducible} to a decision problem $B$ if there exists a polynomial 
time algorithm $\Pi'$ such that, given an instance $x$ of $A$, $\Pi' $ 
produces an instance $z$ of $B$ satisfying $z\in B$ if and only if $x\in A$. 
In such case, we write $A\preceq_{p}B$. 
A problem $A\in$NP is NP-complete if for every $X\in$NP, $X\preceq_{p}A$.  
It is generally believed that NP-complete problems have no polynomial
time algorithms.

Given an undirected graph (not necessarily bipartite) $G=(V,E)$, 
$S\subseteq V$ is a {\em vertex cover} in $G$ if 
for all $(u,v)\in E$ either $u\in S$ or $v \in S$ or both.  
We will be using in our reductions the following decision problems 
associated with the computation of vertex covers in a graph.

\begin{problem}
VERTEX COVER:  Given a graph $G$ and a positive integer $t$ 
does $G$ contain a vertex cover of size at most $t$.
\end{problem}

The above problem is shown to be NP-complete in 
\cite[p. 190]{Gar}.  A variant of this problem referred to by 
the same name and shown to be NP-complete
in \cite[pp. 949--950]{Cor} will be referred to  
here as the following:

\begin{problem}
VERTEX COVER(=):  Given a graph $G$ and a positive integer $t$ 
does $G$ contain a vertex cover of size equal to $t$.
\end{problem}

In the following section we show that both STOPPING DISTANCE
and STOPPING SET are NP-complete by establishing polynomial
time many-one reductions from VERTEX COVER and VERTEX COVER(=) 
respectively to the above problems.

\section{Hardness of STOPPING DISTANCE}
\label{stopping distance}

Let $(G=(V,E),t)$ be an instance of the VERTEX COVER problem.
Let $|V|=n$, $|E|=m$. Excluding trivial cases of the problem
we may  assume $1\leq t\leq n-1$.  We shall make the further
assumption that $G$ is connected.  It is not hard to show that both
VERTEX COVER and VERTEX COVER(=) remain NP-complete 
even when restricted to connected graphs.   

The vertex-edge incidence graph of $G$ is the undirected bipartite graph 
$G'=(L,R,E')$ with $L=V$, $R=E$ and edges $(e,u)$ and $(e,v)$ in $E'$
for each $e=(u,v)\in E$.  Figure 1 shows the vertex-edge incidence
graph for a graph $G$ with $n=4$ and $m=3$.

\begin{figure}
  \begin{center}
     \input{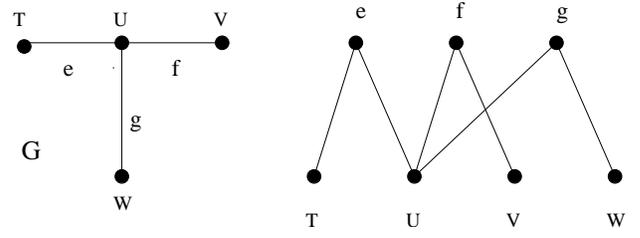}
    \caption{A graph G and its vertex-edge incidence graph}
    \label{figure1}
  \end{center}
\end{figure}

The advantage of assuming that $G$ is connected
arises out of the following lemma:

\begin{lemma}
\label{lemma1}
Let $G'=(L,R,E')$ be the vertex-edge incidence graph of a 
connected graph $G=(V,E)$. Let $S$ be a stopping set in $G'$. 
Then $S=L$.
\end{lemma}
\begin{proof}
Let $L\setminus S\neq \emptyset$.  Then, as $G$ is connected 
there exists $v\in L\setminus S$ and $u\in S$
such that $(u,v)\in E$.  Let $e=(u,v)$.  Then $e\in N(S)$.  
Since $S$ is a stopping set $|N(\{e\})\cap S|\geq 2$.  But
the only neighbours of $e$ in $G'$ are $u$ and $v$.  Hence $v\in S$
contradicting $v\in L\setminus S$.  
\end{proof}

We construct an undirected bipartite graph $G''=(L,R,E'')$ as follows:
$L=\bigcup_{i=0}^{m+1}L_{i}$, $R=\bigcup_{j=0}^{m+1}R_{j}$, where,
$R_{0}=\{z_{1},...,z_{m-1}\}$, $R_{j}=\{u_{j}^{r},u\in V\}$ for
$2\leq j\leq m+1$, $R_1=L_0=E$, $L_{i}=\{u_{i}^{l},u\in V\}$ for
$1\leq i\leq m+1$.  Edges in $G''$ are connected as the following:

\begin{itemize}

\item Connect $u_{i}^{l}\in L_{i}$ to $u_{i}^{r}\in R_{i}$, 
      $2\leq i\leq m+1$.

\item Connect $u_{i}^{l}\in L_{i}$ to $u_{i+1}^{r}\in R_{i+1}$.
      $1\leq i\leq m$

\item For each $e=(u,v)$ in $E$, connect $e\in R_1$ to  
      $u$ and $v$ in $L_1$. 

\item For each $e\in E$ Connect $e\in L_0$ to $e\in R_1$.

\item For the purpose of defining the edges between $R_0$ and $L_0$,
      temporarily re-label vertices in $L_0$ as $e_1,e_2,..e_m$ in 
      some arbitrary way.  Add the edges $(e_i,z_i)$ for $1\leq i\leq m-1$
      and the edges $(e_i,z_{i-1})$ for $2\leq i\leq m$.

\end{itemize}

\begin{figure}
  \begin{center}
    \input{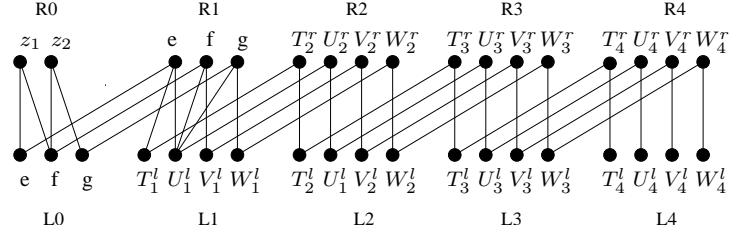}
    \caption{Construction of $G''$}
    \label{figure 2 }
  \end{center}
\end{figure}

The example in figure 3 illustrates the construction of $G''$ for
the graph in figure 1.  The graph $G''$ consists of a copy of the
vertex-edge incidence graph of $G$ (vertex sets $L_{1}$ and $R_{1}$).
Additionally, there are $m$ copies of the vertex set $V$ on the
left ($L_2,L_3,...,L_{m+1}$) and right ($R_2,R_3,...R_{m+1}$).
The connections between $R_{0}$ and $L_{0}$ ensure that any stopping
set in $G''$ containing any one vertex in $L_0$ must contain the
whole of $L_0$.  The vertex  $u_{i}^{r}$ in $R_i$
has neigbours $u_{i-1}^{l}$ and $u_{i}^{l}$ for each $2\leq i\leq m+1$
and each $u\in V$.  This ensures that if a stopping set $S$ in 
$G''$ contains $u_{i}^{l}$ for some $i\in \{1,2,..m+1\}$ then all the
$m+1$ vertices $u_{1}^{l}, u_{2}^{l},....,u_{m+1}^{l}$ 
must be present in $S$.  These observations summarized below
play a crucial role in the arguments that follow. 

\begin{observation}
\label{obs1}
A stopping set $S'$ in $G''$ satisfies $u_{i}^{l}\in S'$ for some
$1\leq i\leq m+1$ if and only if it satisfies
$u_{i}^{l}\in S'$ for {\em every} $1\leq i\leq m+1$.
Moreover either $L_0\subseteq S'$ or $L_0\cap S'=\emptyset$.
\end{observation}

The following two claims establish the connection between vertex covers
in $G$ and stopping sets in $G''$.

\begin{lemma}
\label{IF}
If $G$ contains a vertex cover $S$ of size $t$ 
for some $1\leq t\leq n-1$
then $G''$ contains a stopping set of size $t(m+1)+m$. 
\end{lemma}
\begin{proof}
Consider the set $S'= L_0 \cup \{u_{i}^{l}: u\in S, 1\leq i\leq m+1\}$ 
in $G''$.  Clearly $S'$ has $t(m+1)+m$ elements.
Let $w\in N(S')$.  Then either $w=u_{i}^{r}$ 
for some $u\in S$, $i\in\{2,3,...m+1\}$ or $w\in R_1$ or $w\in R_0$.
In the first case, both $u_{i}^{l}$ and $u_{i-1}^{l}$ are neighbours
of $w$.  If $w\in R_1$, then by construction, $w$ must correspond to
some edge $e=(u,v)$ in $E$. Since $L_0\subseteq S'$,
$e\in L_0$ is a neighbour of $w$.  Since $S$ is a vertex cover in $G$, 
either $u$ or $v$ or both must belong to $S$.  Hence one or both of
$u_{1}^{l}$ and $v_{1}^{l}$ must be a neighbour of $w$ in 
$S'$.  Finally if $w\in R_0$, then both the neighbours of $w$ are
in $L_0$, and therefore in $S'$.  Thus in all cases $w$ has at least 
two neighbours in $S'$.  Consequently $S'$ is a stopping set.  
\end{proof}

We now prove that every stopping set in $G''$ of size less than
$n(m+1)+m$ must correspond to some vertex cover of size $t$ in $G$ 
for some $1\leq t\leq n-1$ and must have size exactly $t(m+1)+m$

\begin{lemma}
\label{ONLYIF}
Let $S'$ be a stopping set in $G''$ of size less than $n(m+1)$. 
Then the following must hold:
\begin{itemize}
  \item $L_0\subseteq S'$, 
  \item $|S'|=t(m+1)+m$ for some $1\leq t\leq n-1$ and 
      $|S'\cap L_i|=t$ for every $1\leq i\leq m+1$
  \item $S=\{u\in V$ : $u_{i}^{l}\in S'$ for some $1\leq i\leq m+1\}$ 
      is a vertex cover of size $t$ in $G$. 
\end{itemize}
\end{lemma}

\begin{proof}
Suppose $L_0$ is not contained in $S'$. Then by Observation~\ref{obs1}
$L_0\cap S'=\emptyset$. Since $S'\neq \emptyset$, There must be
some $u\in V$ and $i\in\{1,2,..,m+1\}$ such that $u_{i}^{l}\in S'$.
By Observation~\ref{obs1} $u_{1}^{l}\in S'$.  Since vertices in 
the set $R_1$  are connected only to $L_1$ and $L_0$, 
every neighbour of $S'$ in $R_1$ must have two neigbours in $S'\cap L_1$
in order for $S'$ to satisfy the conditions of a stopping set.  
In other words, $S'\cap L_1$ 
must be a stopping set in the subgraph of $G''$ induced by the vertices
$L_1\cup R_1$.  Note that this subgraph is the vertex-edge incidence
graph of $G$.  Applying Lemma~\ref{lemma1} we get $S'\cap L_1=L_1$. 
Hence Observation~\ref{obs1} shows that $S'=\bigcup_{i=1}^{m+1}L_i$.  But
in that case $|S'|=n(m+1)$, a contradiction.  Hence $L_0\subseteq S'$ 
and $|L_1\cap S'|<n$. 
Let $|S'\cap L_1|=t$ for some $1\leq t\leq n-1$.  Applying 
Observation~\ref{obs1} once again, $|S'\cap L_i|=t$ for all 
$1\leq i\leq m+1$.  Hence $|S'|=t(m+1)+m$.

To complete the proof of the lemma, it is sufficient to prove that
$S=\{u\in V : u_{1}^{l}\in S'\}$ is a vertex cover of $G$. Since
$L_0\subseteq S'$, $R_1\subseteq N(S')$.  Since every vertex $e$ in $R_1$ 
has only one neighbour in the set $L_0$, 
for $S'$ to satisfy the stopping set condition 
$e$ must have a neighbour in $L_1\cap S'$. 
Then, by construction $\{u\in V: u_{1}^{l}\in S'\}$  must be a vertex 
cover in $G$ as required.  
\end{proof}

As a consequence of Lemma~\ref{IF} and Lemma~\ref{ONLYIF} we have:

\begin{cor}
\label{cor1}
$G$ has a vertex cover of size $t$ if and only if $G''$ 
has a stopping set of size $t(m+1)+m$, $1\leq t\leq n-1$.
Hence $(G,t)\in$ VERTEX COVER(=) if and only if 
$(G'',t(m+1)+m)\in$ STOPPING SET.
\end{cor}

\begin{cor}
\label{cor2}
$G$ has a vertex cover of size at most $t$ if and only if $G''$ 
has a stopping set of size at most $t(m+1)+m$, $t\in\{1,2,..,n-1\}$.
Hence $(G,t)\in$ VERTEX COVER if and only if 
$(G'',t(m+1)+m)\in$ STOPPING DISTANCE.
\end{cor}

We are now ready to prove:

\begin{theorem}
STOPPING DISTANCE and STOPPING SET are NP-complete
\end{theorem}
\begin{proof}
We have proved that $(G,t)\in$ VERTEX COVER if and only
if $(G'',t(m+1)+m)\in$ STOPPING SET. 

Since $G''$ can be constructed
from $G$ in polynomial time ($O(mn)$ time suffices), it 
follows that VERTEX COVER(=) $\preceq_p$ STOPPING SET and
VERTEX COVER $\preceq_p$ STOPPING DISTANCE from
Corollary~\ref{cor1} and Corollary~\ref{cor2} respectively.
It is easy to verify whether a given set of left vertices of a bipartite
graph forms a stopping set in time linear in the size of the
graph. Hence both STOPPING DISTANCE and STOPPING SET belong to
the class NP.   
\end{proof}

As a consequence, we have:

\begin{cor}
\label{Harddist}
Computing stopping distance in a Tanner graph is NP-hard.
\end{cor}

\section{Acknowledgment}
\label{ack}
The authors would like to thank Dr. L. Sunil Chandran for 
helpful discussions.

\end{document}